\documentclass[a4paper,10pt,twoside,superscriptaddress]{cpc-hepnp}
\usepackage{amssymb}
\usepackage{multicol}
\usepackage{graphicx}
\usepackage{booktabs}
\usepackage{amssymb,bm,mathrsfs,bbm,amscd}
\usepackage[tbtags]{amsmath}
\usepackage{lastpage}
\usepackage[colorlinks,linkcolor=blue,anchorcolor=blue,citecolor=blue,dvipdfm]{hyperref}

\begin{document}

\fancyhead[c]{\small Submitted to Chinese Physics C~~~Vol. xx, No. x (2016) xxxxxx}
\fancyfoot[C]{\small 010201-\thepage}

\footnotetext[0]{Received xx June 201x}

\title{Mini-Orange Spectrometer at CIAE\thanks{Supported by National Natural Science
Foundation of China (11305269, 11375267, 11475072, 11405274, 11205068, 11175259) }}

\author{%
      Zheng Yun (Ö£ÔÆ)$^{1)}$\email{zhengyun@ciae.ac.cn}%
\quad Wu Xiao-Guang (ÎâÏþ¹â)$^{2)}$\email{wxg@ciae.ac.cn}%
\quad Li Guang-Sheng (Àî¹ãÉú)
\quad Li Cong-Bo (Àî´Ï²©)\\
\quad He Chuang-Ye (ºØ´´Òµ)
\quad Chen Qi-Ming (³ÂÆôÃ÷)
\quad Zhong Jian (ÖÓ½¡)
\quad Zhou Wen-Kui (ÖÜÎÄ¿ü)\\
\quad Deng Li-Tao (µËÁ¢ÌÎ)
\quad Zhu Bao-Ji (Öì±£¼ª)
}
\maketitle

\address{%
China Institute of Atomic Energy, Beijing 102413, People's Republic of China\\
}

\begin{abstract}
A Mini-Orange spectrometer used for in-beam measurements of internal conversion electrons, which consists of a Si(Li) detector and different sets of SmO$_5$ permanent magnets for filtering and transporting the conversion electrons to the Si(Li) detector, has been developed at China Institute of Atomic Energy. The working principle and configuration of the Mini-Orange spectrometer are described. The performance of the setup is illustrated by measured singles conversion electron spectra using the Mini-Orange spectrometer.
\end{abstract}



\footnotetext[0]{\hspace*{-3mm}\raisebox{0.3ex}{$\scriptstyle\copyright$}2013
Chinese Physical Society and the Institute of High Energy Physics of the Chinese Academy of Sciences and the Institute of Modern Physics of the Chinese Academy of Sciences and IOP Publishing Ltd}%

\begin{multicols}{2}

\section{Introduction}

There exist two elementary processes of deexcitation in an excited nucleus. The nucleus may deexcite through the $\gamma$-rays emission or by means of the internal conversion electrons emission, by which the excitation energy transfer directly to electrons near the nucleus and then these electrons fly away from the atom. The internal conversion electrons and $\gamma$-rays compete against each other in deexciting an excited nucleus. The internal conversion coefficient $\alpha$ is defined as the ratio of the conversion electron transition probability and the $\gamma$-ray transition probability. The internal conversion process depends much on the atomic number $Z$, the transition energy and the transition multipolarity, $EL$ and $ML$. In the case of medium and heavy nuclei or high multipolarity transitions or low transition energies, $\alpha$ increases rapidly and internal conversion becomes the favored process. In fact, the conversion coefficients, $\alpha_K$ and $\alpha_L$, can be used to determine transition multipolarities and then to assign spins and parities, since they contain information on the electric or magnetic character of the transitions. Moreover, internal conversion electron spectroscopy is the only way to detect 0$\rightarrow$0 transitions. Because the $\gamma$ decay is not allowed due to the spin of a $\gamma$-ray itself is $1\hbar$. Therefore, internal conversion electron spectroscopy paly an important role in the study of nuclear structure.

In order to carry out in-beam conversion electron spectroscopy at an accelerator, an efficient method need to be used to suppress the high background from $\delta$ electrons, $\beta^+$ electrons, and $\gamma$ rays. A new type of spectrometer for conversion electron consisting of a Si detector and a filter of permanent magnets around a central absorber of lead or tungsten has been created by Klinken J V et al. in 1972~\cite{VanKlinken19721}. This new type of Mini-Orange spectrometer (MOS)~\cite{VanKlinken19721,VANKLINKEN1975427,VANKLINKEN1978433,Ditzel1996428} meet the requirements for in-beam measuring conversion electrons due to its advantages of wide energy range, high transmission efficiency, high energy resolution, and easy operation. The MOS has been constructed for a long time and used to measure conversion electrons~\cite{Aengenvoort1998,PhysRevC53652,Gassmann2001181,PhysRevC77024314,Regis2009466,PhysRevLett103012501,Rudigier201089}. Before the present work, however, there is no any MOS has been built and used in nuclear spectroscopy experiments in China. Thus, a new MOS has been developed at China Institute of Atomic Energy (CIAE) for in-beam studies of conversion electrons. In combination with a $\gamma$-ray detector array, e$^-$-$\gamma$-coincidence measurements can be performed.

In this paper, we report the development of MOS at CIAE. In Section 2, the principle and configuration of the MOS are detailed illustrated. In Section 3, the transmission curves measured by using the method of Ref.~\cite{Guerro201432} is briefly described. The test experiment and results are presented in Section 4. A summary is given in Section 5.

\section{The mini-orange spectrometer}

The MOS combine the excellent energy and time resolution of a cooled Si(Li) detector (300 mm$^{2}$ area and 3 mm thick) with the high efficiency and selectivity of a magnetic Mini-Orange filter (MOF). A view of the MOS is shown schematically in Fig~\ref{fig1}. The MOF works similar to an optical lens. The internal conversion electrons within a certain energy range are focused onto the Si(Li) detector by the MOF as a function of the object distance $g$ between the target and the MOF and the image distance $b$ between the MOF and the Si(Li) detector. For fixed electron energies the MOF works according to the well-known optical relation for the focal length $f$
\begin{equation}
\label{one}
\frac{1}{f}=\frac{1}{g}+\frac{1}{b}
\end{equation}
The image distance $b$ can be adjusted by a retractable bellows, and the object distance $g$ can be adjusted by a rotatable MOF as can be seen in Fig.~\ref{fig1}. The large background of low-energy delta electrons ($\delta$) are suppressed by selecting an optimized transmission. This filter also deflects the positive beta electrons ($\beta^+$) to the outer side of the detector.

\begin{center}
\includegraphics[width=7cm]{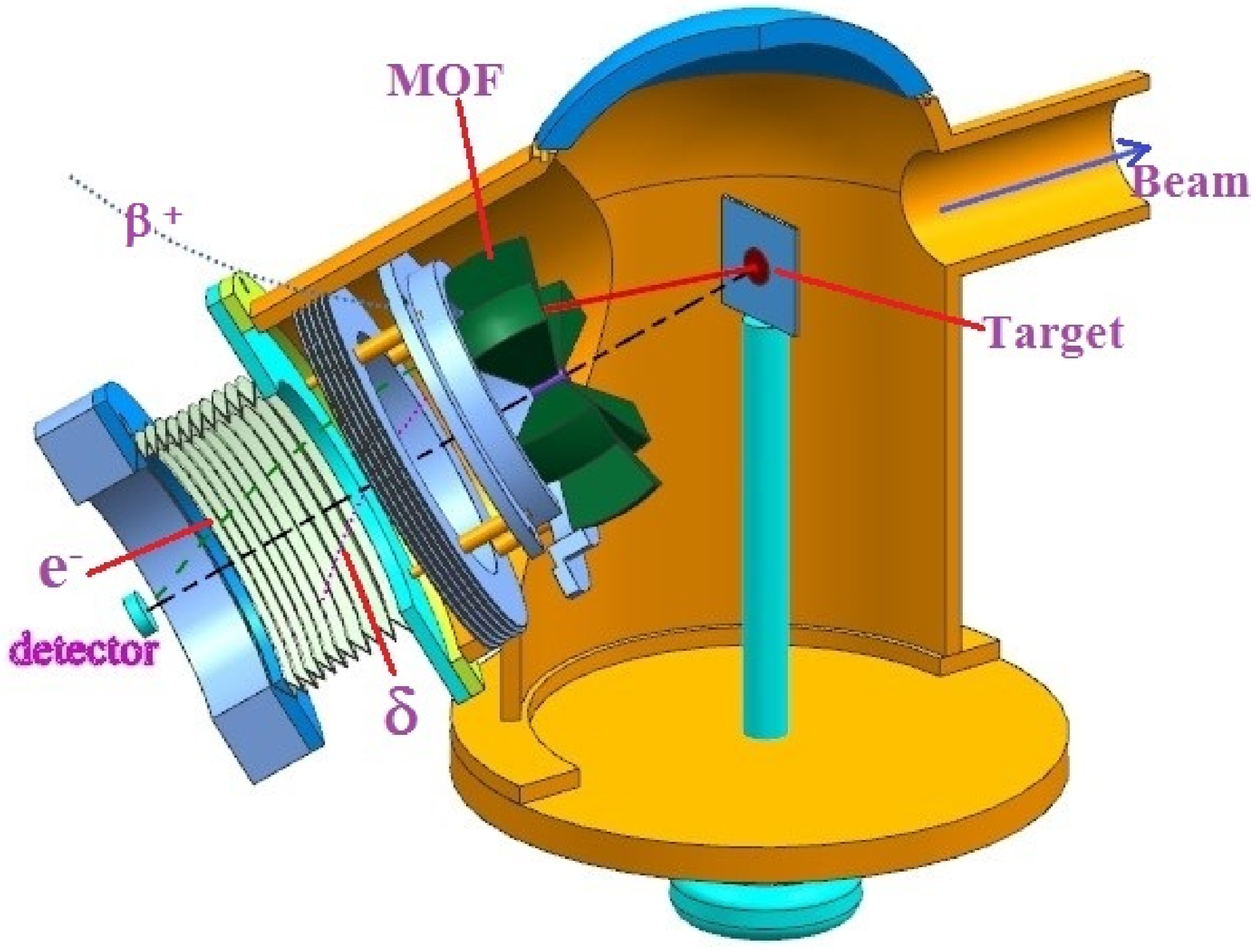}
\figcaption{\label{fig1} (Color online) A schematic view of the Mini-Orange Spectrometer. }
\end{center}

Fig.~\ref{fig2} shows a MOF consisted of three wedge-shaped permanent magnets, which are arranged symmetrically around a cylindrical Pb absorber. The outer radius of the magnet wedges is 4 cm. The angle of the wedges is $5^\circ$ in $\phi$ direction and $70^\circ$ in $\theta$ direction symmetrically around $90^\circ$. The magnets are glued onto an aluminium holder, which is then screwed onto a metallic ring of 7 cm inner and 10.8 cm outer diameter, respectively. On the outer edge of the metallic ring is screw thread, which can be rotated in the target pipe (see Fig.~\ref{fig1}). The lead absorber is 30 mm long and 12 mm in diameter. It is fixed in the centre of the wedges by a notch. $\gamma$-rays originating from the target were blocked by the Pb absorber. Moreover, the Pb absorber was covered by a cadmium cap to absorb X-rays produced in the Pb absorber. This configuration is intended for use at lower energies, while other three similar filters with thicker magnets could be used for higher energies. The compact size of the MOF allows the MOS to be easily integrated into a large array of $\gamma$-ray detectors. The MOS installed at a beam line of HI-13 tandem accelerator at CIAE is shown in Fig.~\ref{fig3}.

\begin{center}
\includegraphics[width=6cm]{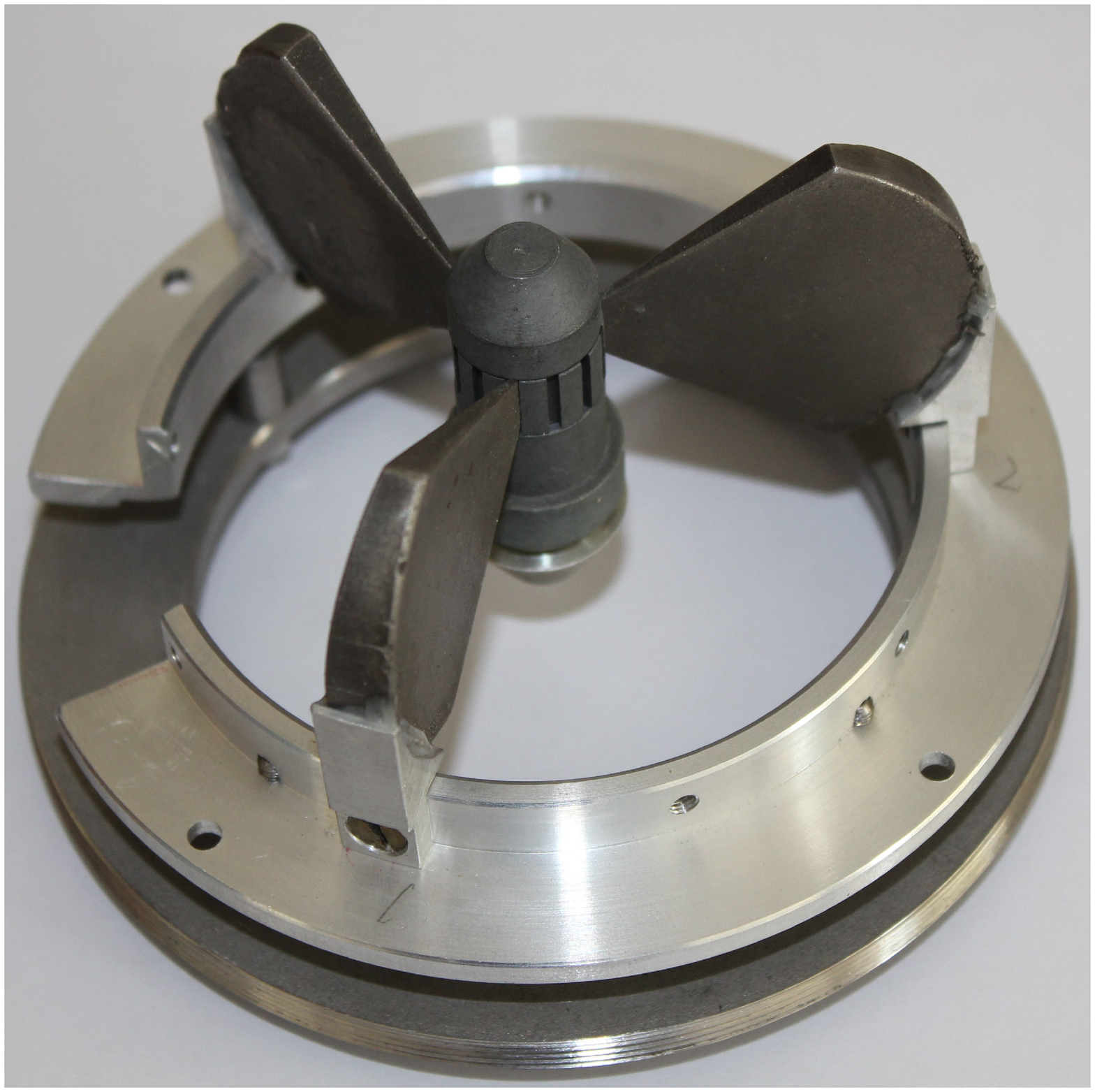}
\figcaption{\label{fig2} (Color online) View of the Mini-Orange filter consisting of 3 wedge-shaped SmCo$_5$ permanent magnets arranged around a cylindrical Pb absorber. The 5 mm thick aluminium cap on top of the absorber reduces X-rays produced by electrons scattered in the lead. }
\end{center}

The wedge-shaped permanent magnets are made of rare-earth materials ceramic SmCo$_5$. Inside the gaps between the magnets the typical toroidal field deflects conversion electrons emitted from a source or target towards the Si(Li) detector. The preferred field strengths range usually from 0.2 to 2 or more kGauss~\cite{Ditzel1996428} and can be selected by choice of the thickness, strength and the number of the magnets (three in Fig.~\ref{fig2}).

\begin{center}
\includegraphics[width=7cm]{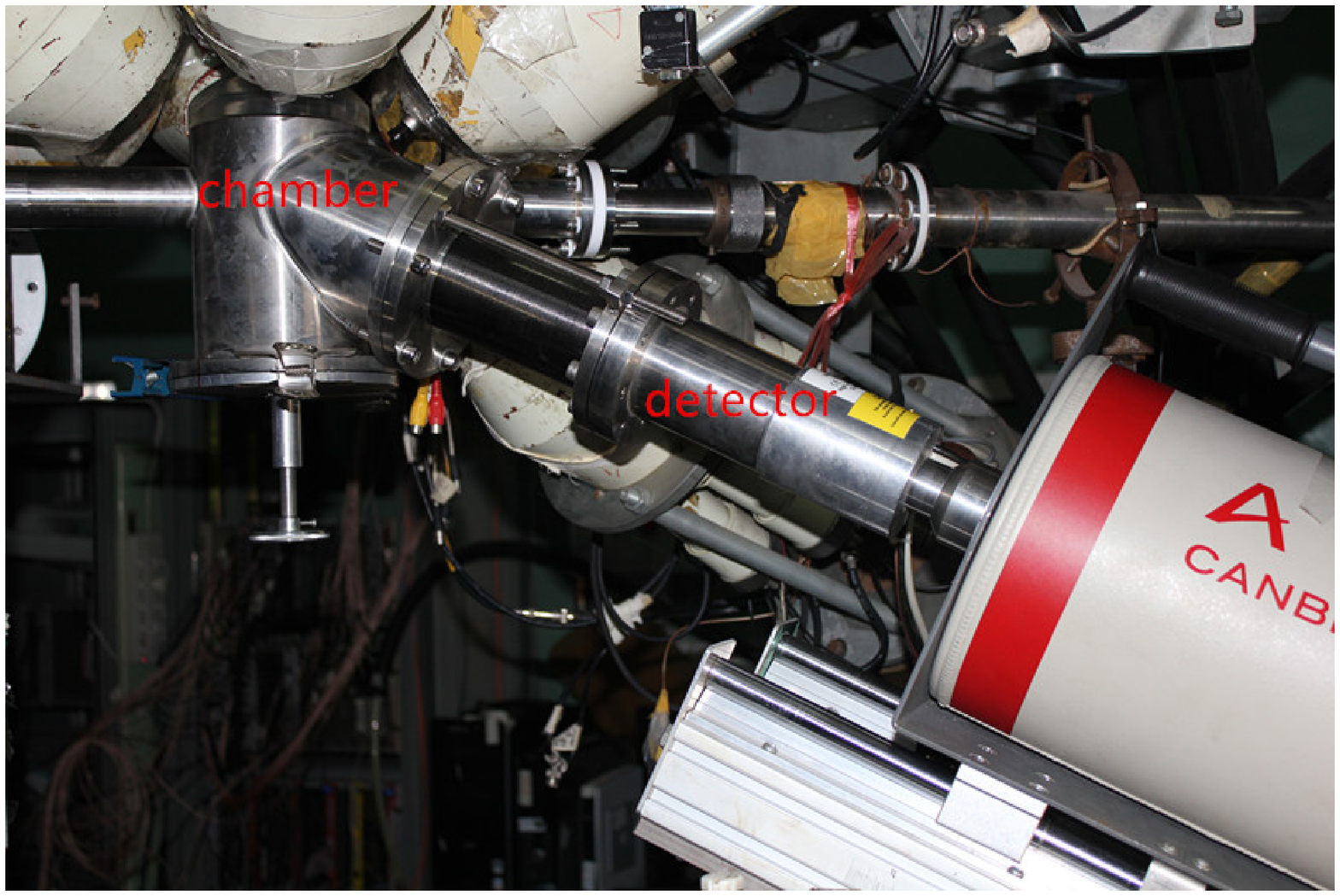}
\figcaption{\label{fig3} (Color online) View of the Mini-Orange Spectrometer at a beam line of HI-13 tandem accelerator at CIAE.}
\end{center}

\section{The MOS transmission curves}

The function of MOF is to filter and transport the conversion electrons to the Si(Li) detector. The transmission efficiency of the MOS is therefore strongly energy dependent. The position and the width of the transmission window depend on the type of MOF and on the distances of the MOF to the target and Si(Li) detector. The transmission $T$ of a MOS is defined as the number of electrons with energy $E$ detected by the detector with the MOS $N_{W}(E)$ compared to the number of electrons with the same energy detected without the MOS $N_{WO}(E)$.
\begin{equation}
\label{two}
T(E)=\frac{N_{W}(E)}{N_{WO}(E)}
\end{equation}
where the label ¡°wo¡± correspond to the case without MOS and ¡°w¡± correspond to the case with MOS.

In the present configuration the distances $g$ and $b$ of the MOF to the target and Si(Li) detector were 60 mm and 145 mm, respectively. The transmission curves have been measured by using the method of Ref.~\cite{Guerro201432}. The continuum energy spectra of $\beta^{-}$ source $^{90}$Sr were measured and were corrected for electron backscattering by using discrete sources $^{152}$Eu. $\beta^-$ spectra of $^{90}$Sr measured with and without MOS, respectively, are shown in Fig.~\ref{fig4} (a). Fig.~\ref{fig4} (b) shows the measured transmission curve deduced from the ratio of $N^{exp}_{W}(E)/N^{exp}_{WO}(E)$. The resulting transmission curve (green) is shown in Fig.~\ref{fig5}. Five red dots are the experimental transmission points of $^{152}$Eu. As can be seen in Fig.~\ref{fig5}, the largest value of transmission efficiency is at about 200 keV in the range of energies from 25 to 350 keV.

\begin{center}
\includegraphics[width=8cm]{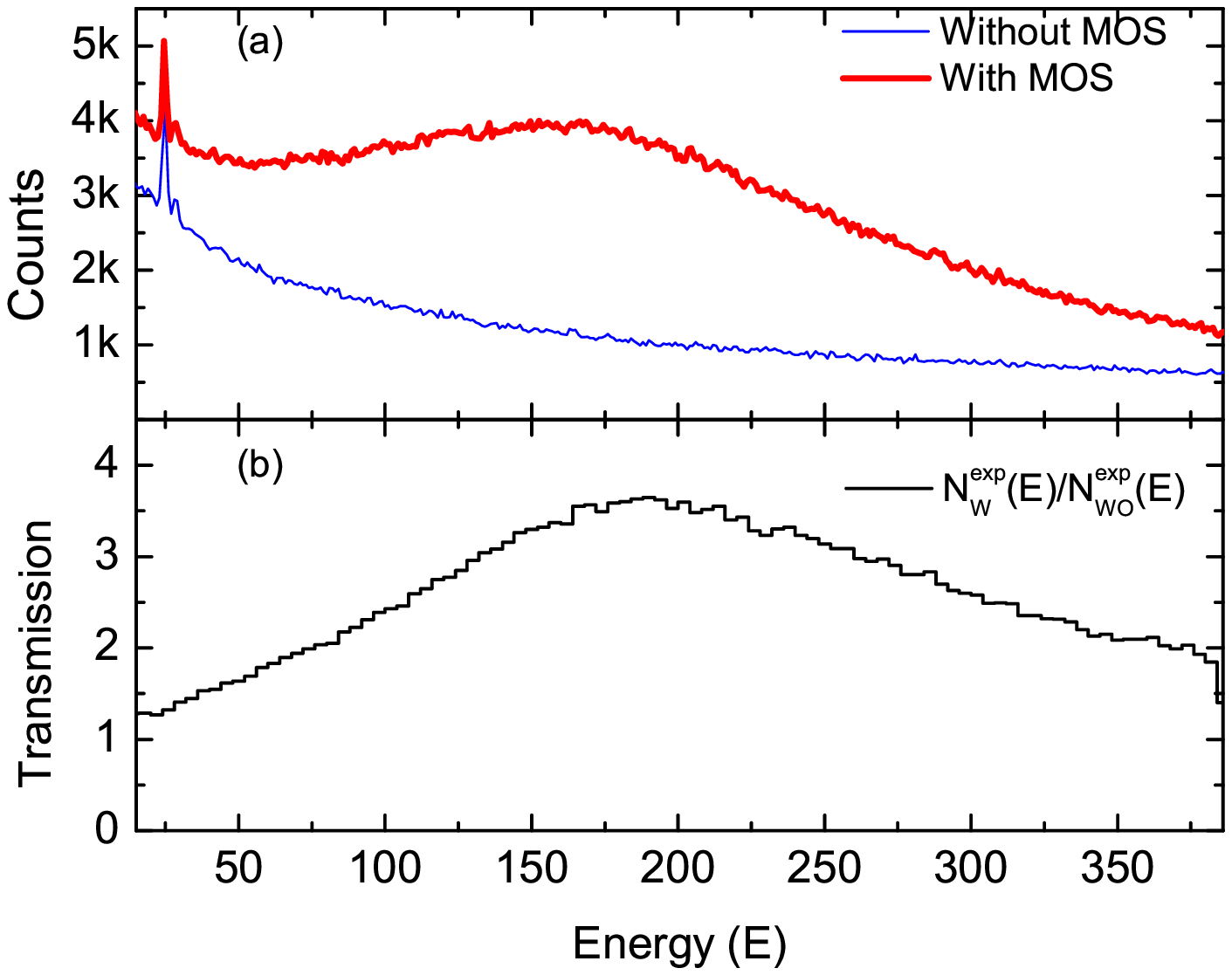}
\figcaption{\label{fig4} (Color online) (a) $\beta^-$ spectra of $^{90}$Sr measured with and without MOS, respectively, for the configuration g=60 mm and b=145 mm with three magnets. (b) The curve corresponds to the ratio $N^{exp}_{W}(E)/N^{exp}_{WO}(E)$ of the measured $^{90}$Sr spectra.}
\end{center}

\begin{center}
\includegraphics[width=8cm]{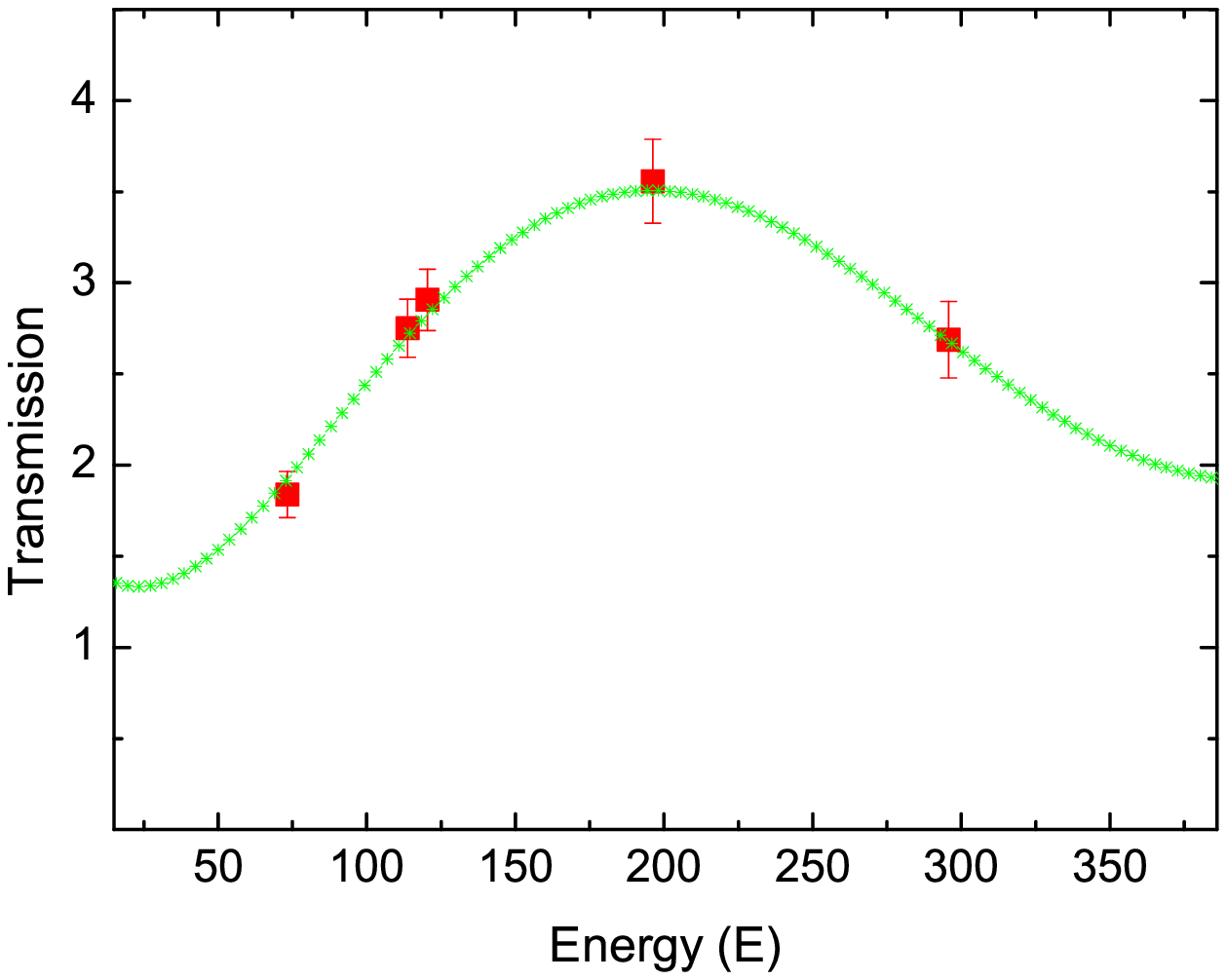}
\figcaption{\label{fig5} (Color online) Transmission curve for the MOS for the configuration g=60 mm and b=145 mm with three magnets. The transmission
curve (green) was obtained using the method of Ref.~\cite{Guerro201432}. The red dots are the experimental transmission points of $^{152}$Eu.}
\end{center}

\section{Testing of the MOS}

The major interest of this work is to investigate the feasibility of in-beam internal conversion electron measurements using the MOS. For this purpose, the conversion electrons of the the high-spin isomeric state in $^{93}$Mo ($T_{1/2}=6.85$h, $E_x=2.425$MeV, $I^{\pi}=21/2^+$) was measured. This isomeric state was first identified by Kundu et al.~\cite{Kundu1950}. The $K/(L+M)$ ratio of conversion electron suggests an $E4$ character of the 263-keV transition deexciting the isomeric state~\cite{Alburger1953}. Furthermore, some other studies~\cite{PhysRevC8315,PhysRevC132166,PhysRevC15390,PhysRevC232252,Fukuchi2005,PhysRevC80034306} have been carried out to research the nucleus.

Excited states in $^{93}$Mo were populated via the heavy-ion fusion-evaporation reaction $^{90}$Zr$(\alpha$, $n)^{93}$Mo. The $\alpha$ beam, with an energy of 30 MeV, was delivered by the HI-13 tandem accelerator at the China Institute of Atomic Energy. The target was a 0.43-mg/cm$^{2}$-thick isotopically enriched $^{90}$Zr metallic foil with a 9.5-mg/cm$^{2}$ Au backing to stop the recoiling nuclei. The conversion electrons were recorded by a MOS adjusting to an optimized transmission. The MOS was placed at $135^\circ$ with respect to the beam direction. The absolute energy calibration of MOS was performed with conversion electron lines from $^{133}$Ba, $^{152}$Eu, and $^{137}$Cs sources; the energy resolution of the Si(Li) detector was 1.7 keV at 624 keV. The $\gamma$ rays emitted from the evaporation residues were also detected with a multi detector array consisting of 9 BGO-Compton-suppressed HPGe detectors, whose energy resolutions were about 2.0-2.5 keV at 1.33 MeV, and two planar HPGe detectors with energy resolutions of 0.6-0.7 keV at 121.78 keV. The HPGe detectors were kept at 42$^{\circ}$, 90$^{\circ}$, 140$^{\circ}$, and 153$^{\circ}$ with respect to the beam direction. All HPGe detectors were calibrated for energy and efficiency using the standard energy calibration $\gamma$ lines from the decay of $^{133}$Ba and $^{152}$Eu radioactive sources. The gate width of prompt $\gamma$ timing was set to 500 ns. A total of $3.0\times10^6$ $\gamma-e$ and $\gamma-\gamma$ coincidence data were accumulated in event-by-event mode. After energy calibration and gain matching for different detectors, the recorded $\gamma-e$ and $\gamma-\gamma$ coincidence events were sorted into two-dimensional $E_{e}-E_{\gamma}$ and $E_{\gamma}-E_{\gamma}$ symmetric matrixes, and then analyzed using the software package RADWARE~\cite{Radford1995}. Some new experimental results obtained from the coincidence data will be discussed elsewhere. In the present paper, we just focus on the singles spectrum of internal conversion electrons.

The conversion electron spectra distinctly show L lines in connection with the K lines so that K/L ratios can be determined. The analysing method of the present singles conversion electron spectra can be briefly summarized as follows: (1) The background spectrum will be subtracted from the original singles conversion electron spectrum; (2) The spectrum after background subtraction will be calibrated with the transmission curve; (3) The counts of the K and L transition lines will be registered, then the K/L ratios will be deduced. The in- and off- beam singles conversion electron spectra from the high-spin isomeric state in $^{93}$Mo measured by the MOS are presented in Figs.~\ref{fig6} and ~\ref{fig7}.

\begin{center}
\includegraphics[width=8cm,height=8cm]{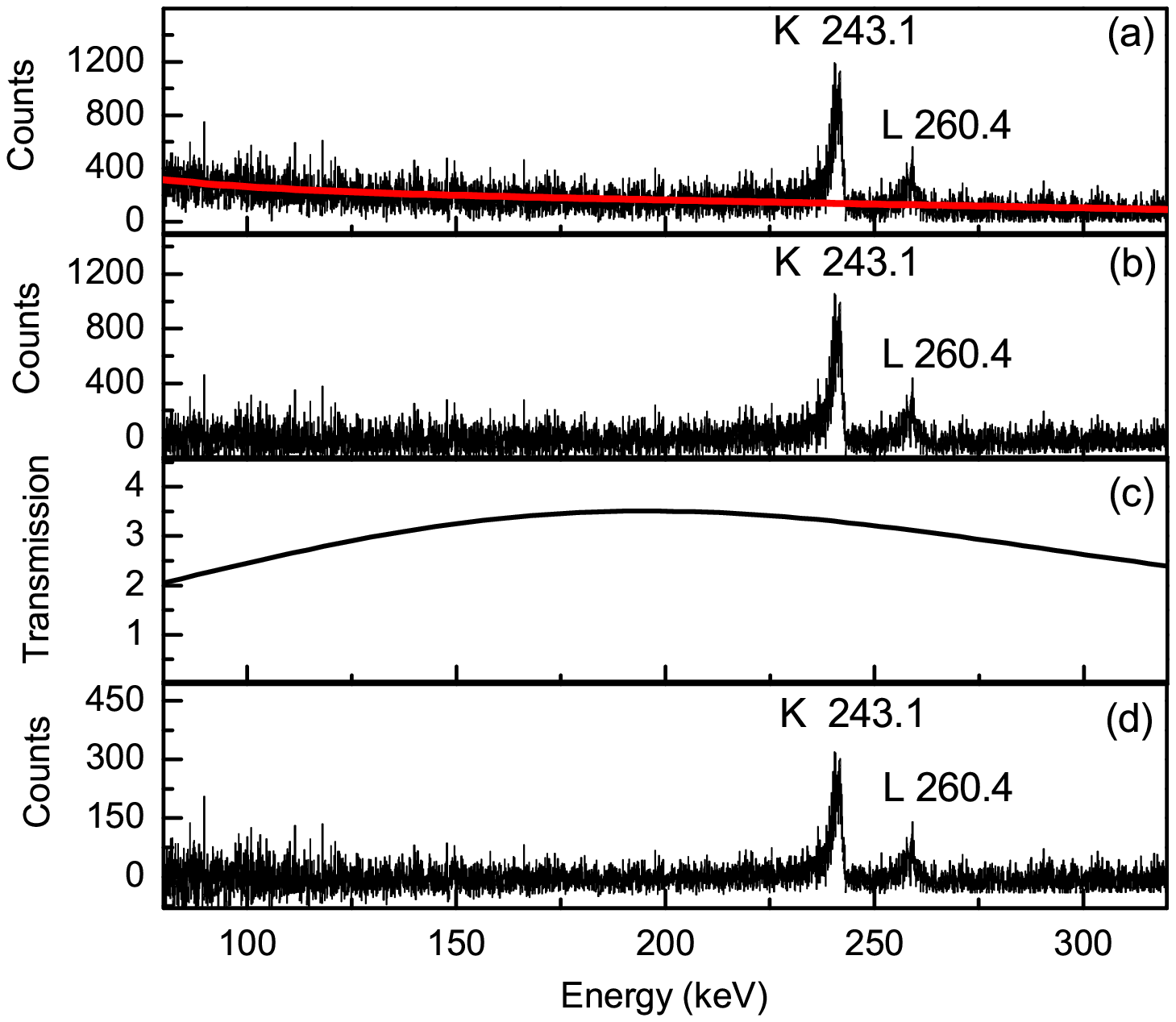}
\figcaption{\label{fig6} (Color online) (a) The measured off-beam singles conversion electron spectrum emitted from the $21/2^+$ isomeric state in $^{93}$Mo. The red curve corresponds to the fitted background spectrum. (b) The off-beam singles conversion electron spectrum after background subtraction. (c) The transmission curve. (d) The off-beam singles conversion electron spectrum after background subtraction and efficiency calibration with the transmission curve. The figure shows the spectrum of K and L conversion peaks correspond to the 263-keV transition.}
\end{center}

\begin{center}
\includegraphics[width=8cm,height=8cm]{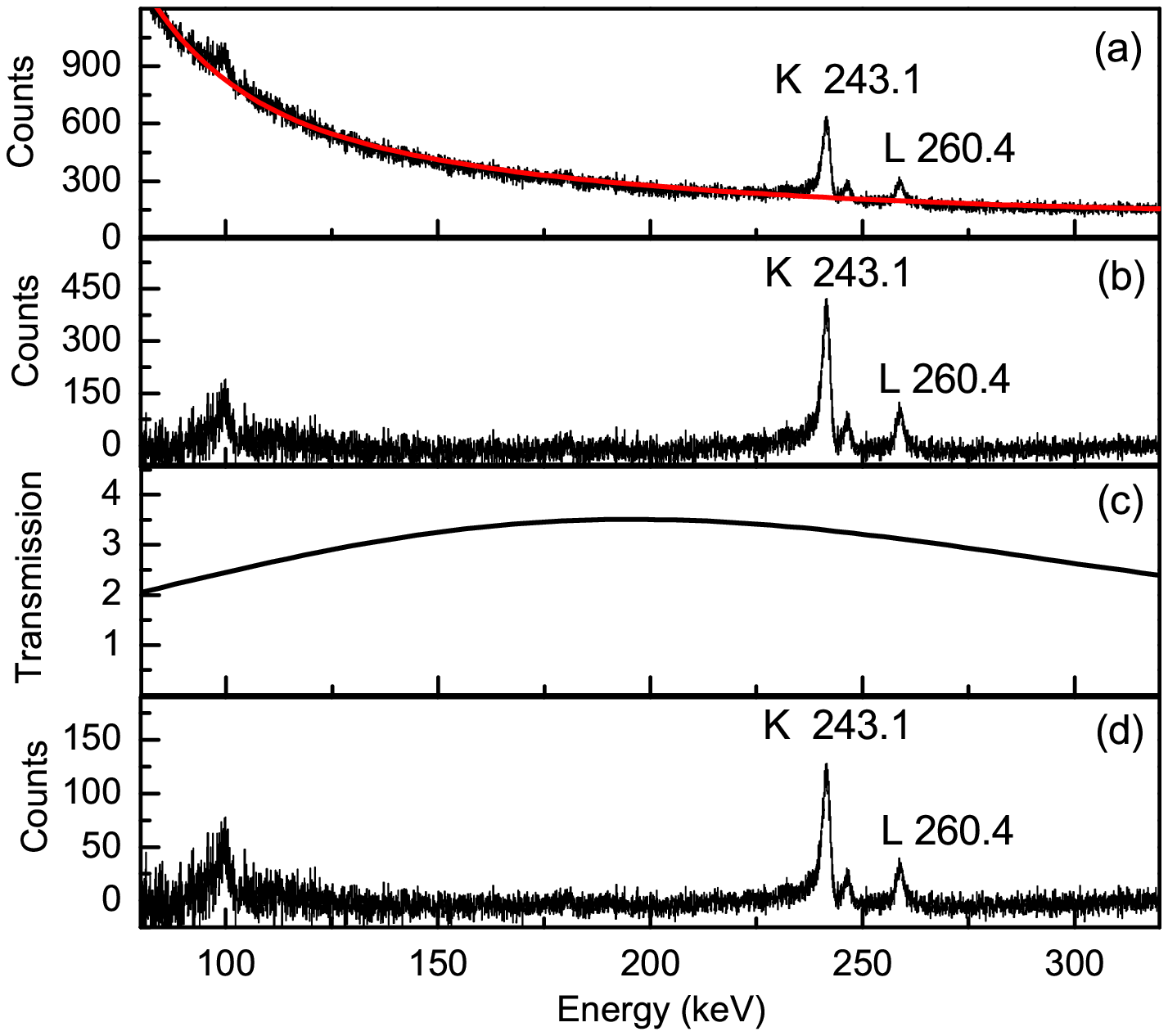}
\figcaption{\label{fig7} (Color online) (a) The measured in-beam singles conversion electron spectrum emitted from the $21/2^+$ isomeric state in $^{93}$Mo. The red curve corresponds to the fitted background spectrum. (b) The in-beam singles conversion electron spectrum after background subtraction. (c) The transmission curve. (d) The in-beam singles conversion electron spectrum after background subtraction and efficiency calibration with the transmission curve. The figure shows the spectrum of K and L conversion peaks correspond to the 263-keV transition.}
\end{center}

In the original off-beam singles spectrum (Fig.~\ref{fig6}(a)) the peaks for K and L conversion electrons, which correspond to the 263-keV $\gamma$ transition of $^{93}$Mo, can be clearly observed. The red curve corresponds to the fitted background spectrum. In Fig.~\ref{fig6}(b), the off-beam singles conversion electron spectrum emitted from the $21/2^+$ isomeric state in $^{93}$Mo after background subtraction is given. The transmission curve is given in Fig.~\ref{fig6}(c). The off-beam singles spectrum of the conversion electrons after background subtraction and efficiency calibration with the transmission curve is shown in Fig.~\ref{fig6}(d). The K and L line at transition energies of 243.1 keV and 260.4 keV are clearly visible. The binding energy of the K and L electrons are 19.9 and 2.6 keV, respectively. In the K transition line 18118(1631) counts could be detected, while in the L transition line 5039(907) counts were registered. The resulting K/L ratio of 3.60(72) is consistent with the theoretical K/L ratio of 3.47~\cite{Hager1968} and corresponds to the $E4$ character suggested in Ref.~\cite{Alburger1953}.

The illustrations of Fig.~\ref{fig7}(a)-(d) are similar to those of Fig.~\ref{fig6}. Fig.~\ref{fig7}(d) shows the in-beam singles conversion electron spectrum after background subtraction and efficiency calibration with the transmission curve. The K and L transition lines with energies of 243.1 keV and 260.4 ke correspond to the 263-keV transition can clearly be identified. 5143(463) and 1479(222) counts were measured in the K and L lines, respectively. The resulting K/L ratio of 3.48(61) is also consistent with the $E4$ character suggested in Ref.~\cite{Alburger1953} and the theoretical K/L ratio of 3.47~\cite{Hager1968}.

\section{Summary}

A new compact Mini-Orange Spectrometer consisting of a cooled Si(Li) detector and the high efficiency and selectivity of a magnetic Mini-Orange filter has been developed for the use of in-beam spectroscopy experiments at CIAE. The setup was tested using the heavy-ion fusion-evaporation reaction $^{90}$Zr$(\alpha$, $n)^{93}$Mo. The background from $\delta$ electrons, $\beta^+$ electrons, and $\gamma$ rays is highly suppressed. The K and L lines at transition energies of 243.1 keV and 260.4 keV corresponding to the 263-keV transition are clearly visible so that K/L ratios can be determined. The in- and off- beam resulting K/L ratios of 3.48(61) and 3.60(72) , respectively, are consistent with the theoretical K/L ratio of 3.47~\cite{Hager1968} and corresponds to the $E4$ character suggested in Ref.~\cite{Alburger1953}. The new MOS has turned out to be a useful instrument for the in-beam measurements of conversion electrons, and provided a new method for identification of low-lying transitions and for assignment of spin and parity.
\\

\acknowledgments{The authors would like to thank the crew of the HI-13 tandem accelerator at the China Institute of Atomic Energy for steady operation of the accelerator and for preparing the target. This work is partially supported by the National Natural Science Foundation of China under Contract No. 11305269, No. 11375267, No. 11475072, No. 11405274, No. 11205068, and No. 11175259.}

\end{multicols}
\vspace{-1mm}
\centerline{\rule{80mm}{0.1pt}}
\vspace{2mm}
\begin{multicols}{2}

\end{multicols}

\clearpage
\end{document}